\documentclass[10pt]{iopart}

\usepackage{iopams}
\usepackage{color}
\usepackage{graphicx}% Include figure files
\usepackage{dcolumn}% Align table columns on decimal point
\usepackage{bm}% bold math
\usepackage[english]{babel}
\usepackage{multirow}
\DeclareUnicodeCharacter{0301}{\'{e}}
\usepackage{siunitx}
\usepackage{xcolor}

\begin{document}

\title[]{Skyrmions: A review on materials perspective for future electronic devices}% Force line breaks 
\author{Vineet Kumar Sharma$^1$, Alana Okullo$^1$, Jalen Garner$^1$, Cheng Peng$^2$, Rajan Plumley$^2$, Adrian Feiguin$^3$, Chunjing Jia$^4$, Josh Turner$^2$, A. Bansil$^3$, Sugata Chowdhury$^1$}
\address{$^1$Department of Physics and Astronomy, Howard University, Washington DC, 20059}
\address{$^2$SLAC Stanford National Accelerator Laboratory, 2575 Sand Hill Rd, Menlo Park, CA 94025}
\address{$^3$Physics Department, Northeastern University, Boston, MA 02115}
\address{$^4$Department of Physics, University of Florida, 2001 Museum Rd. Gainesville, FL 32611-8440}
\ead{ar.bansil@northeastern.edu}
\ead{sugata.chowdhury@howard.edu}

\begin{abstract}
Recent years have witnessed an enormous rise in research interest in magnetic skyrmions owing to their capability to improve over contemporary spintronic devices. An overview of the various magnetic interactions responsible for the formation of skyrmion together with distinct noncentrosymmetric and centrosymmetric skyrmion candidates is given in this review article. The magnetic interactions known as Dzyaloshinskii–Moriya interactions (DMI) have been extensively studied over the years to better understand the mechanism of skyrmions in chiral magnets that have larger skyrmion sizes. Because of their low skyrmion size, the centrosymmetric/frustrated magnets are dwelling to skyrmions controlled by long-range interactions such as the Ruderman–Kittel–Kasuya–Yosida interaction (RKKY), which may be useful in the development of high-density memory devices. To lay a solid foundation for the magnetic interactions involved in skyrmion formations and many other special physical properties, more research in the field of centrosymmetric skyrmions is required. Apart from studying candidates with low skyrmion sizes, one of the main goals for the future is to better understand the dynamics of skyrmion using polarized magnons, which has the potential to be extremely beneficial for spintronic applications.
\end{abstract}

\maketitle
\ioptwocol
%\tableofcontents

\section{Introduction}

Electronic devices and their components have been constructed in ever-tinier spaces since the 1950s, when the transistor was discovered, to maximize computing speed and memory storage volume~\cite{RevModPhys.71.S336}. Miniaturization efforts at the current rate will soon become unsustainable, and new techniques appropriate for the demands of today's technologically advanced society will be needed for any further advancement of computing power. Research and development of devices that depend on the spin degree of freedom of electrons rather than just their charge are the main goals of these efforts. We refer to this field of study as spintronics~\cite{JOSHI20161503}. The Skyrmion, named after the theorist Tony Skyrme, has been a worthy candidate of interest in recent years. Skyrme made significant contributions, one of which was the idea of the topological number. In his model of nonlinear theory for interacting pions, he postulated the concept of skyrmions, which are pion field arrangements that resemble particles~\cite{Skyrme:1958vn}. As a result, in 1989~\cite{luo2021skyrmion}, skyrmions in liquid material were theoretically predicted. A diffraction study conducted in 2009~\cite{muhlbauer2009skyrmion} confirmed the existence of the magnetic skyrmion, which was then observed in 2010 through electron microscopy~\cite{yu2010real}. Ever since skyrmions have been thoroughly studied in many different systems. 

In general, skyrmions are quasiparticles characterized by noncollinear, magnetic swirl-like spin textures that are determined by topologically stable field configurations. Skyrmions can form as individual particles or in crystal lattices of magnetic materials like helices or hexagonal structures~\cite{Fert2017}. Due to the implications of skyrmion magnetic, non-trivial, real-world topology, these quasiparticles have attracted a lot of attention in the field of computational device research. They have been extensively studied and identified in both centrosymmetric and noncentrosymmetric lattice structures. High stability in the ferromagnetic environment at the nanoscale favors the transportation of information by electric spin, electric charge, or their combination~\cite{luo2021skyrmion}. Other consequences include the ability of electrons to undergo discontinuous density changes to become ferromagnetic states, as well as anomalous and topological Hall effects~\cite{nagaosa2010anomalous}. The goal of more research on skyrmions is to potentially solve the complexities of contemporary computers. Research attempts to further reduce computational device sizes while also increasing system complexity, and speed, and lowering energy consumption. Applications of skyrmion research include the creation of memory devices, transistors, skyrmion racetracks, diodes, synapses, and neurons, among other things. Nevertheless, skyrmions experience subsequent current-driven motion, which reduces their utility in spintronic devices. Skyrmion Hall Effect can result in data loss or pinning even though it induces electric-current excitation~\cite{luo2021skyrmion}.

\begin{figure*}[!ht]
    \center
    \includegraphics[width=\textwidth]{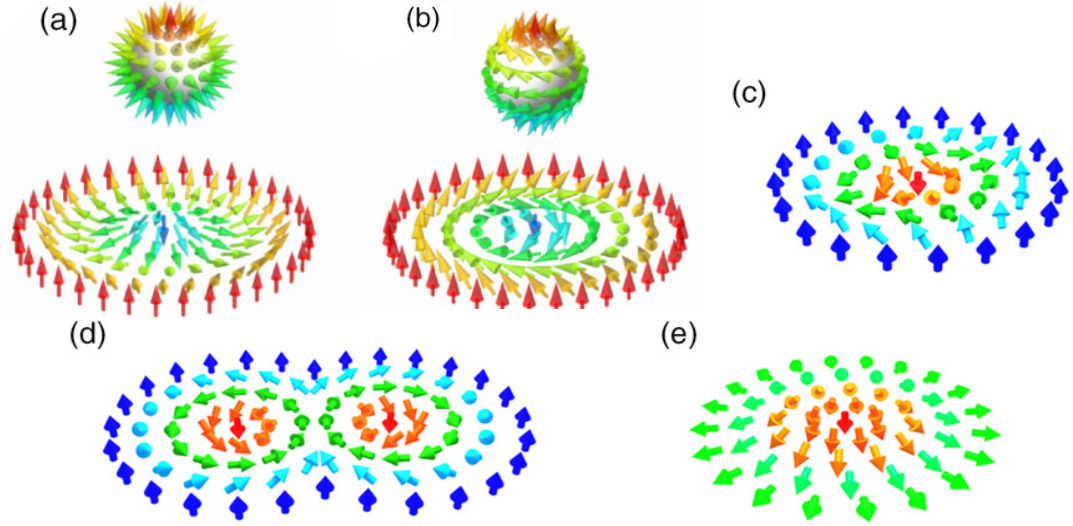}
    \caption{Visualization of different types of two-dimensional topological spin textures: (a) N\'eel-type skyrmion, (b) Bloch-type skyrmion, (c) antiskyrmion, (d) biskyrmion, (e) meron. Reproduced from reference~\cite{ma2022recent} }
  \end{figure*}

Now, to offer a more intuitive description of the Skyrmion and its properties, we will refer to Fig 1. Skyrmions are stabilized via dipole-dipole interactions~\cite{PhysRevLett.105.197202}, Dzyaloshinskii–Moriya interaction (DMI)~\cite{Ham_Pradipto_Yakushiji_Kim_Rhim_Nakamura_Shiota_Kim_Ono_2021}, Ruderman–Kittel–Kasuya–Yosida interaction (RKKY)~\cite{Yambe_Hayami_2021}, and geometric frustration~\cite{PhysRevB.94.144405}, caused by conflicting interatomic forces of the material. DMI, the antisymmetric spin-exchange interaction, can turn the spin alignment to create the fixed swirl-like texture characteristic of the skyrmion. This interaction emerges from relativistic spin-orbit coupling as well as the lack of structural inversion symmetry. It is always present in magnets with noncentrosymmetric crystal lattice structures~\cite{dzyaloshinsky1958thermodynamic, moriya1960anisotropic}. These skyrmions have fixed vorticity and helicity. Three types of skyrmions have been categorized to date: (i) N\'eel-type in chiral magnets [Fig. 1(a)], (ii) Bloch-type in polar magnets [Fig. 1(b)], and (iii) Anti-skyrmions [Fig. 1(c)]~\cite{Chen_2017}. There are two types of skyrmionic hosts: bulk materials that intrinsically have the capability of forming skyrmions, and heterostructured synthetic materials with DMIs at the interface~\cite{GOBEL20211, tokura2020magnetic}. Skyrmions are most commonly found in the interfaces of noncentrosymmetric magnets. However, in structures that have a center of symmetry, skyrmions can be generated using geometric frustration coupled with RKKY interactions, a long-range interaction~\cite{10.1063/1.5062168}. Because of this interaction, stabilized skyrmions can be ultra-small with the magnitude of a few nanometers.\\
Magnetic skyrmions have successfully been generated and erased in thin heterostructure films and bulk-polar lattice magnets~\cite{GOBEL20211, tokura2020magnetic}. Magnetic systems have yielded predictions and measurements of the anomalous Hall effect in several coplanar Kagome magnets without net magnetization and the crystal Hall effect, for which the arrangement of non-magnetic atoms in the crystal breaks a set of time reversal and spatial symmetries~\cite{PhysRevResearch.2.033112}. Their influence on conventional skyrmions is not yet well explored. In 2017, Antisymmetric skyrmions were first discovered above room temperature in tetragonal Heusler compounds, characterized by having the boundary walls of alternating N\'eel and Bloch-type skyrmions~\cite{Nayak_Kumar}. A challenge for crystal-structure control in thin-film growth still lies ahead. Large-scale growth of the thin film with a single-chirality crystal domain is required. Since the number of known bulk materials hosting N\'eel-type skyrmions is still limited, the detailed skyrmion nature in polar magnets remains to be explored. Recent findings claim the centrosymmetric systems as a host to skyrmions governed by frustrated interactions~\cite{kurumaji2019skyrmion}. In centrosymmetric materials, the appearance of skyrmions is an advantage over noncentrosymmetric materials, as their helicity and polarity are non-fixed. These aspects can be controlled as an additional degree of freedom, which is better applicable to the quantum computing industry, especially for the development of the skyrmion qubit~\cite{tran2022electrical}. Determining the magnetic interactions is the challenge after identifying centrosymmetric skyrmion materials.

\begin{figure*}[!ht]
    \center
    \includegraphics[width=\textwidth]{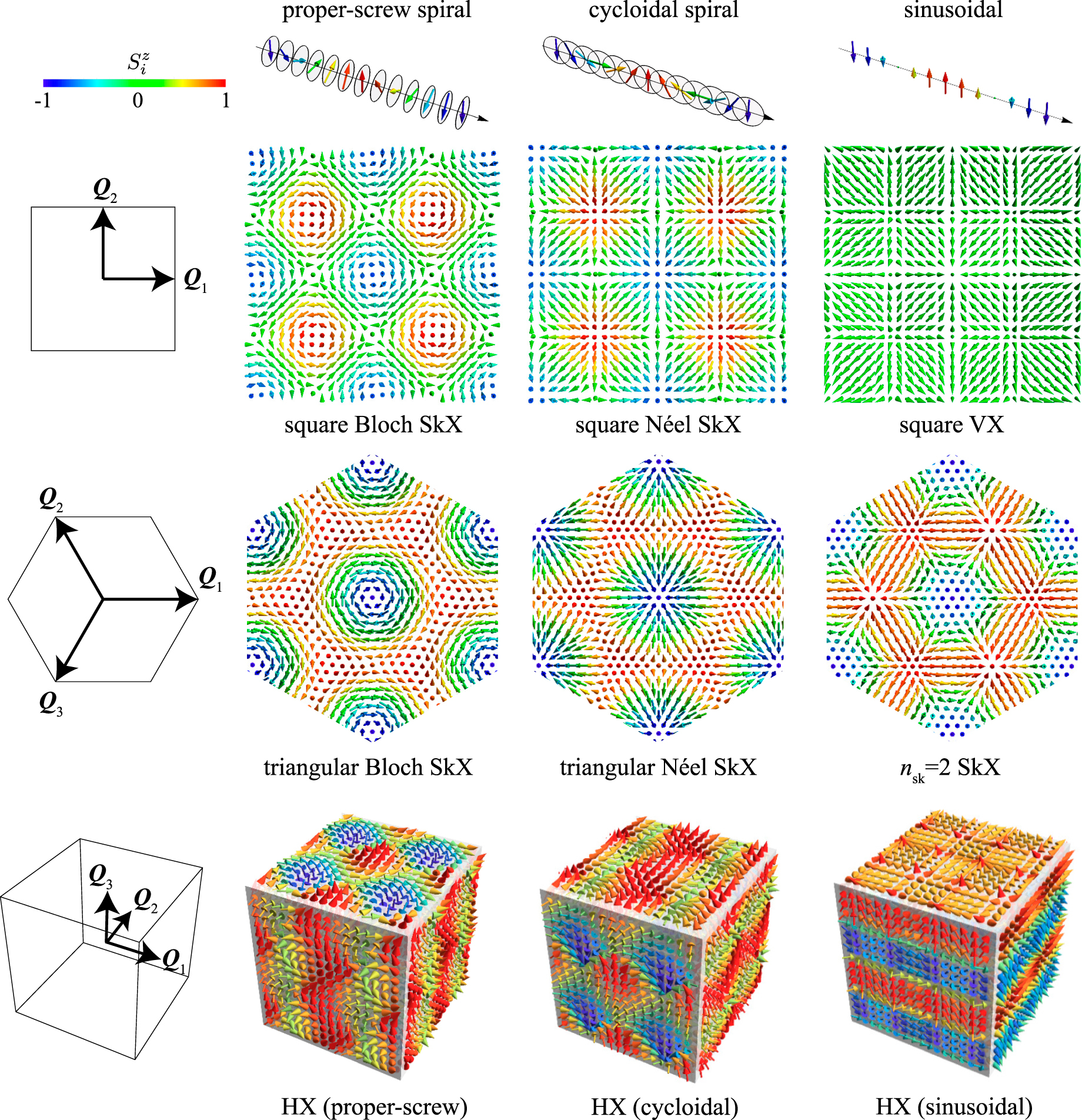}
    \caption{Diagram of topological spin textures as seen by the superposition of proper-screw spirals, cycloidal spirals, and sinusoidal waves, where the color of the arrows corresponds to the z-component direction. Each row corresponds to double-q square lattices, triple-q triangular lattices, and 3D triple-q cube lattices. SkX, VX, and HX each represent the skyrmion, vortex, and hedgehog crystal. Reproduced from reference~\cite{hayami2021topological}}
  \end{figure*}

In this review, we would like to demonstrate the current state of noncentrosymmetric and centrosymmetric skyrmion materials. This will be done in the following manner; section II illustrates a thorough description of the theory of topological spin formations, offering a view into the mechanisms by which they are formed and additionally providing context on the different types of spin formations that are related to skyrmions: antiskyrmions, merons and hedgehogs. Next, the current state for the various noncentrosymmetric and centrosymmetric skyrmions candidate materials will be addressed. In doing this, we plan to provide ample information on the mechanics by which the skyrmion state arises within these materials. Further, the technological aspects of skyrmion phases are addressed. In conclusion, we will provide an outlook for those seeking to study chiral and non-chiral magnets and discuss any challenges that future studies will face.

\section {Theoretical Discussion: Topological Spin Textures and magnetic exchange interactions}

As aforementioned in the introduction section, magnetic skyrmions, amongst other topological spin textures, present a host of desirable properties and are realized in a wide range of bulk materials and thin film structures. In an attempt to provide readers with the necessary background, we will define noncollinear and noncoplanar formations in a manner that is both flexible and precise, to facilitate better understand of the theory underlying these structures. When we think about electrons, they have two principal characters: charge and spin. We can generally categorize the electron’s spin into collinear and noncollinear spin patterns. Furthermore, noncollinear spin structures can be subdivided as either noncoplanar (chiral) or coplanar structures. Noncollinear materials are host to skyrmions, other spin textures, and emergent phenomena like the topological Hall effect, the multiferroic effect, anomalous Hall effect and spin-polarized currents in antiferromagnets, Weyl fermions, and the spin Hall effect under certain circumstances~\cite{qin2020noncollinear}. 
In recent years, noncollinear and noncoplanar spin textures have attracted much attention in condensed matter physics due to these various emergent properties. These phenomena arise due to the magnetic order emerging from different possible spin-spin interactions~\cite{hayami2021topological}. These orders are characterized by order parameters such as the spin vector chirality, a vector product of spins 
$\textbf{D}_{ij}\cdot (\textbf{S}_i\times \textbf{S}_j)$ and the triple scalar product, \(\textbf{S}_i \cdot (\textbf{S}_j \times \textbf{S}_k)\), with $i,j,k$ are the site coordinates that forming spin textures, known as the spin scalar chirality~\cite{batista2016frustration, xiao2010berry, nagaosa2010anomalous}. Various spin textures have been theorized to occur by taking the superposition of multiple spin density waves, known as multiple-$Q$ magnetic states~\cite{batista2016frustration, shapiro1979multiple}. There can be multiple variations of the skyrmion spin textures, as seen in Fig.1, each representing different inner mechanisms explaining their spin patterns. The three most common of these are the Bloch-type (Fig 1. (b))~\cite{muhlbauer2009skyrmion,yu2010real,seki2012observation}, N\'eel-type [Fig. 1(a)]~\cite{PhysRevLett.119.237201}, and antiskyrmion[Fig. 1(c)], differing from each other by their spin modulations along the radial direction. There are also different non-skyrmion spin textures that have been discovered and subdivided by their unique topological charges~\cite{nagaosa2013topological, braun2012topological}. Here we use $Q$ to represent the topological charge. For two-dimensional spin textures, these are the meron and anti-meron of topological charge $Q = \pm \frac{1}{2}$, Fig. 1(e), and the bi-skyrmion with topological charge $Q = -2$, Fig. 1(d). Likewise, in three-dimensions there is the hedgehog and anti hedgehog, with topological charge $Q = +1$ and $Q = -1$, respectively. (Fig. 2 bottom row). These spin textures are generally characterized by their topological charge $Q$ and one should have knowledge its theoretical background.  To clarify further the theory behind these spin formations, the topological charge can be defined as

\begin{eqnarray}
\hspace{1cm}
Q = \frac{1}{4\pi}\int_{A} \vec{n}(\vec{r}) \cdot [{\partial_i\, \vec{n}(\vec{r})} \times {\partial_j\,\vec{n}(\vec{r})}] d\vec{r}
\end{eqnarray}

In this case, $\vec{n}(\vec{r})$ equals the spin moment direction and is defined as 

\begin{equation}
\hspace{3cm} \vec{n}(\vec{r}) = \frac{\vec{m}(\vec{r})}{|\vec{m}(\vec{r})|}
\end{equation}

Here, $\vec{m}(\vec{r})$ are magnetization fields of spin textures in spherical coordinates and defined as $\vec{m}(\vec{r}) = m(\sin\Theta(\vec{r})\cos\Phi(\vec{r}), \sin\Theta(\vec{r})\sin\Phi(\vec{r}), \cos\Theta(\vec{r}))$, with $\vec{r} = (r\cos(\phi), r\sin(\phi))$ in polar coordinates.
The integration range is a region $A$ that contains the entire spin texture, and the indices $i, j$ represent independent directions in the chosen reference system. For two-dimensional radially symmetric spin textures, $A$ is to be defined as a circular disk, and $i, j$, can be in either cartesian $(x, y)$ or polar $(r, \theta)$ coordinates. Likewise, for three-dimensional and radially symmetric cases, $A$ is a sphere, and $(i, j)$ are represented by spherical coordinates $(\theta, \phi)$ as the polar and azimuthal angles, respectively. Additionally, $Q$ overall is representative of the number of times that the sum of all the spin moment directions $\vec{n}(\vec{r})$ wraps around a unit sphere. 

Inputting $\vec{n}(\vec{r})$ in terms of $\vec{m}(\vec{r})$ into Eq.(2.1), with $R$ as the spin texture radius, $Q$ can be redefined as:
\begin{equation}
\hspace{1cm}
 Q = -\frac{1}{4\pi} [\cos\Theta(r)]_{r=0}^{r=R} \, [\Phi(\phi)]_{\phi=0}^{\phi=2\pi}   
\end{equation}

The former part on the right-hand side of equation 2.3 is linked to the magnetic moment orientation with respect to a magnetic field ($\vec{H}$). In skyrmions, the core magnetic moments are antiparallel to $\vec{H}$ where, $\Theta(r)=\Theta(0)=\pi$ and those moments on the periphery orient parallel where $\Theta(r)=\Theta(R)=0$. This combination makes $[\cos\Theta(r)]_{r=0}^{r=R} = 2$. Latter part $\omega = [\Phi(\phi)]_{\phi=0}^{\phi=2\pi}$ in equation 2.3 defines the vorticity, where $\omega$ is the integer number of azimuthal rotations of moments $\vec{n}(\vec{r})$ along a closed path that encircles the spin texture center. $\omega$ further defines $Q$ as $Q = -\omega$. Furthermore, $\Phi$ is equal to:
$\Phi = \omega\phi + \gamma$,
where, $\gamma$ is the helicity, and this relation arises if the azimuthal angle $\Phi$ of the moment monotonically changes with the azimuthal direction of $\phi$ and $r$, $\Phi$ positions~\cite{tokura2020magnetic}. Skyrmions are thus defined by $(Q, \omega, \gamma)$ and based upon these quantities various types of skyrmions are tabulated in Table:1. In general terms, Bloch-type skyrmions feature moment rotations that are in a plane perpendicular to the radial direction, N\'eel-type moments rotate in a plane along the radial direction and antiskyrmion moments alternate between the Bloch and N\'eel-type rotations. 
\\

\begin{table}[h!]
\centering
\setlength{\arrayrulewidth}{0.5mm}
\setlength{\tabcolsep}{12pt}
\renewcommand{\arraystretch}{1.5}
\begin{tabular}{ |p{1cm}|p{1cm}|p{1cm}|p{1.7cm}|  }

\hline
\multicolumn{4}{|c|}{Topological Charge, Vorticity, and Helicity}\\
\hline

($Q,\omega,\gamma$)&\multicolumn{3}{|c|}{Skyrmion Type}\\
\hline
&Bloch&N\'eel&Antiskyrmion \\
\hline
$Q$&$-1$&$-1$&$1$\\
\hline
$\omega$&$1$&$-1$&$-1$\\
\hline
$\gamma$&$\pi/2$&$0$&$-\pi/2$\\
\hline
\end{tabular}

\caption{The skyrmion types with their respective Topological Charge ($Q$), Vorticity ($\omega$), \& Helicity ($\gamma$).}
\end{table}

Skyrmions are typically found forming regular arrangements. The skyrmion crystal (SkX) phase can be described by its spin density wave or hybridized triple-$q$ state~\cite{hayami2021locking}. The topological charge of these different textures cannot be changed from one $Q$ to another, leading to topological protection to these spin formations. For a triangular lattice SkX, each of these three $q$-vectors $(Q_1, Q_2, Q_3)$ are perpendicular to the applied magnetic field direction (for example, $z$), forming a $120^{\circ}$ angle with each other in a two-dimensional plane. The selection of a single spin moment vortex spiral of this kind in a hexagonal SkX would lead to a Bloch-type skyrmion with topological charge ($Q = -1$)~\cite{hayami2021topological}. If the spin moments $\vec{n}(\vec{r})$ are restricted on a circular plane perpendicular to the single $q$-vector, the resultant spin formation is cycloidal or N\'eel-type. If this is given a triple-$q$ state, the resultant formation is a N\'eel-type SkX. Similarly, when given a double-$q$ state with each $q$-vector orthogonal to each other, along the (100) and (010) planes, the resultant formation is a meron crystal of $Q = -\frac{1}{2}$ and anti meron of $Q = \frac{1}{2}$ for a tetragonal lattice. Because of this meron- anti-meron interaction, the total skyrmion number is $Q = 0$. Thus, it carries no net topological charge. In the case where the triple-$q$ vectors are spread over [100], [010], [001] directions, the resultant structure is the hedgehog or anti-hedgehog lattice. The key to understanding the skyrmion is developing a sense of its topological-particle-like properties as well as the protection that is induced by this topology~\cite{hayami2021topological}.                                                                                                                                    

Microscopic mechanisms are required to twist the direction of the electronic spins to form skyrmions and related noncollinear spin textures. Two primary forms of magnetic interactions serve as the foundation for these twisting mechanisms, which are amenable to the stable skyrmions that have been detected thus far. These broadly discussed magnetic interactions are: (i) the Dzyaloshiskii-Moriya Interaction (DMI): the antisymmetric exchange interaction due to spin-orbit coupling and (ii) long-ranged interactions mediated by itinerant electrons, resulting in magnetic frustration. The Dzyaloshinskii-Moriya interaction (DMI) is an antisymmetric exchange interaction originated by spin-orbit coupling in materials without inversion symmetry. These features in noncentrosymmetric magnets enable them suitable for skyrmion candidates stabilized via DM interactions dictated by a specific twisting rule dominated by the crystal symmetry of the target material and results in the tilting of adjacent spin moments~\cite{dzyaloshinsky1958thermodynamic}. Three types of skyrmions, Bloch-type, N\'eel-type, and antiskyrmion, have been observed in crystals exhibiting chiral, polar, and $D_{2d} / S_4$ symmetries, respectively. The strength of DM interaction is evaluated in terms of the DMI parameter, D$_{ij}$ between two neighboring spins at `i' and `j' sites, respectively. The influence of DM interaction on the formation of noncollinear spin textures is explained by a phenomenological theory using a continuum model of the magnetization field $\vec{m}(\vec{r})$~\cite{chikazumi1997physics}. The stable $\vec{m}(\vec{r})$ in its minimized energy state is contributed by all magnetic interactions, among which $w_{DM}$ is DM interaction energy density. This $w_{DM}$ is position-dependent and can be described by Lifshitz invariants as mentioned in literature~\cite{tokura2020magnetic} for various crystallographic symmetry. These expressions describing $w_{DM}$ in terms of vorticity ($\omega$) and helicity ($\gamma$) for a particular crystallographic symmetry are signatures of the fixed skyrmion state corresponding to single species characteristics in noncentrosymmetric materials. This characteristic requires that other than DM interaction, some additional effects including Zeeman energy due to the external magnetic ﬁeld, dipole interaction energy, and magnetic anisotropy are also necessary to stabilize skyrmions.

In the quest to discover the skyrmion phase in noncentrosymmetric magnets, it was also observed that thermal fluctuations effects become vital for skyrmion formation in narrow temperature regions in these magnets~\cite{doi:10.1126/science.1166767}. Further analysis introduced the mode-coupling mechanism as a key player in understanding the skyrmion phase in systems with multiple helical states. This mode-coupling mechanism can not be only facilitated via thermal fluctuations but long-range exchange interactions such as Ruderman-Kittel-Kasuya-Yosida (RKKY) interaction also support this~\cite{heinze2011spontaneous, okubo2012multiple, leonov2015multiply, hayami2017effective, grytsiuk2020topological, kronlein2018magnetic, romming2018competition, brinker2019chiral}. Long-range magnetic interactions occurring between sites, further than just the nearest neighbor, have resulted in magnetic frustration causing different types of spin textures. Geometric frustration in materials with triangular or Kagome lattices can also mediate in multi-$q$ formation. Unlike DMI, which requires other additional effects such as Zeeman energy, magnetostatic interaction, and magnetic anisotropy to stabilize the skyrmion state, the long-range interactions can lead to skyrmion formation in centrosymmetric materials. More research in this direction could lead to the discovery of more promising materials that host skyrmions, such as nano-sized skyrmions and other topological spin structures that can have a variety of emergent properties of their own, especially in light of the recent discovery of this effect. The noticeable facts about these long0range interactions are 1) these can stabilize isolated Skyrmions~\cite{bezvershenko2018stabilization} and 2) Skyrmions formed via these mechanisms are typical of the order of nanoscale driven by correlations between the length scale of magnetic interactions and the modulation period~\cite{hayami2021topological}. The long-range nature of the Rudderman-Kittel-Kasuya-Yosida (RKKY) interaction leads to magnetic frustration or competing exchange interactions in itinerant magnets. These effects were observed in form of topological spin crystals in Fe/Ir interfaces~\cite{heinze2011spontaneous}, MnGe~\cite{kanazawa2011large}, Gd-based compounds~\cite{kurumaji2019skyrmion, hirschberger2019skyrmion, khanh2020nanometric} and SrFeO$_3$~\cite{ishiwata2020emergent}. The illustration of these interactions has been reviewed here in the forms of a few noncentrosymmetric skyrmions in section III and some centrosymmetric skyrmions in section IV.

\section{Noncentrosymmetric skyrmions}
Chiral magnets exhibit the noncentrosymmetric crystal structure and helical magnetic ordering, enabling them suitable candidates for topological spin textures like skyrmions driven by DMI interaction. Intermetallic compounds like MnSi and FeGe crystallized in space group P2$_1$3~\cite{muhlbauer2009skyrmion, yu2011near, yu2010real, munzer2010skyrmion} and iron langasite
Ba$_3$NbFe$_3$Si$_2$O$_{14}$, crystallizing in space group P3$_1$2~\cite{marty2008single, stock2011helical} set a benchmark for a new class of magnetic materials called chiral magnets. The journey of discovering various chiral materials is stimulated by novel concepts of magnetic exchange interactions present in them responsible for topological spin textures such as skyrmions which were later found to host many spintronic applications. In this section, we would like to discuss various aspects of noncentrosymmetric skyrmions in different chiral materials.

\subsection{The family of B20 compounds}
The family of B20 compounds consists of transition metal and group 14 elements (Si, Ge, or Sn) with a composition ratio of 1:1, and are crystallized into the chiral cubic lattice structure with space group P2$_1$3[Fig. 3(a)]. The absence of inversion symmetry is ensured by an opposite spiral stacking of transition metal or group 14 atoms, which enables DMI in conjunction with spin-orbit coupling (SOC). Because of the canted neighboring spin moments, the combined effect of DM and SOC leads to the formation of helical and skyrmion spin textures with long modulation periods ranging from 10-100 nm. Hedgehog lattice-like three-dimensional spin crystals are found in a variety of B20 magnetic materials~\cite{yu2010real, kanazawa2017noncentrosymmetric, seidel2016topological, yu2011near, shibata2013towards, grigoriev2013chiral, munzer2010skyrmion, pfleiderer2010skyrmion} except their Ge-based alloys~\cite{kanazawa2012possible, fujishiro2019topological} with ultrashort period of 5 nm. The different magnetic structures identified in B20 materials are shown in Fig. 3(c). In addition to large DMI, these ultrashort periods require the involvement of other important magnetic interactions. The formation of helical and skyrmion states in these chiral cubic magnets is primarily influenced by three magnetic exchange interactions: ferromagnetic exchange interaction, DM interaction, and magnetocrystalline anisotropy~\cite{bak1980theory, nakanishi1980origin}. These magnetic interactions' wide energy scale variations cause changes in magnetic states as a function of temperature and magnetic field. The long-period helical structure at zero magnetic fields is determined by the competition between ferromagnetic and DM interactions. The direction of the modulated helical is determined by the weakest magnetocrystalline anisotropy. Such conditions give rise to multidomain states made up of helices propagating in various crystallographic directions. These helical structures align into a single conical structure that propagates along the direction of the external magnetic field when a magnetic field stronger than the critical field is applied. Ferromagnetic transition results from further enhancement of the magnetic field. Using small-angle neutron scattering (SANS), the two-dimensional triangular lattice of Skyrmions in MnSi was observed~\cite{kusaka1976ultrasonic, date1977electron, kadowaki1982magnetization} in different measurements, much earlier than when Muhlbauer et al. claimed their formation for the first time\cite{muhlbauer2019magnetic}. The observed magnetic structure is a superposition of three helical structures created by a modulated magnetic wave vector diffracting an unpolarized neutron beam into two independent wave vectors. The 2D skyrmion lattice is produced by these helical structures, which are angled toward one another at 120$^{\circ}$ in a plane perpendicular to the magnetic field. Following up on these observations, other research groups used Lorentz Transmission Electron Microscopy~\cite{yu2010real} to show the intricate topological spin arrangements within the skyrmion. The achievement of the skyrmion state in MnSi and Fe$_{0.5}$Co$_{0.5}$Si has been demonstrated to serve as a model for skyrmion states in numerous other B20 materials. Certain physical quantities such as specific heat and magnetic susceptibility that exhibit distinct anomalies are useful parameters for the detection of skyrmion states~\cite{bauer2012magnetic, bauer2013specific}. 

\begin{figure}[ht!]
\centering
\includegraphics[width=0.4\textwidth]{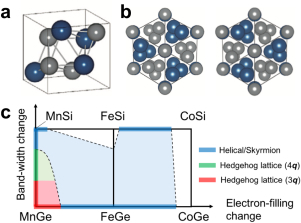}
\caption{The unit cell of B20-type crystal structure. Blue and gray spheres represent transition-metal and group-14-element atoms, respectively. (b) The crystal structures of the two enantiomers viewed from the [111] crystal axis. (c) Overview of the magnetic states in B20-type mixed crystals with 3d-transition-metal elements. Thick color lines indicate the experimentally observed results, and color shades represent the anticipated magnetic states. Red, green, and blue colors denote 3q-hedgehog lattice, 4q-hedgehog lattice, and helical/skyrmion states, respectively. Reproduced from reference~\cite{tokura2020magnetic}.}
\end{figure}

In addition, the magnetic ordering temperature (T$_N$) achieved by the chemical substitution method involving Ge elements on Si sites was found to be higher than that of B20 silicide materials. Due to the substitution of Ge on the Si site, two distinct types of hedgehog lattices (HL) were formed: the triple-q (3q) HL in MnSi$_{1-x}$Ge$_x$ (0.7$\le$x$\le$1) and the quadruple-q (4q) HL in MnSi$_{1-x}$Ge$_x$ (0.3$\le$x $\le$0.6). Three orthogonal q propagating independently along [100] axes characterize 3q-HL, while four q propagating in the apical direction of a regular tetrahedron corresponding to [111] axes characterize 4q-HL. These HLs display magnetocrystalline energy and the competing DM interaction, just like B20 silicides. The underlying mechanism of origin of such spin textures may be interesting for the development of next-generation electronic devices and is crucial for the realization of topological spin textures in chiral magnets.

\subsection{Co-Zn-Mn alloys}
Co-Zn-Mn alloys are those chiral magnets that exhibit stable skyrmion at and above room temperature~\cite{tokunaga2015new}, a key factor for realizing skyrmion-based device applications. The origin of different kinds of skyrmions can occur on these magnets due to tunable parameters such as disorder, magnetic anisotropy, and magnetic frustration~\cite{karube2016robust, karube2018disordered, karube2017skyrmion, yu2018transformation}. The stable topological spin states in these materials will now be our main concern. Co-Zn-Mn alloys are cubic chiral lattices of the $\beta$-Mn type with space group P4$_1$32/P4$_3$32. They can be understood as two possible combinations of oppositely handed atoms that are joined by a mirror operation[Fig. 4(a)].
These alloys give rise to geometric frustration of antiferromagnetic interactions because of Co atoms are positioned at crystallographic site 8c with three-fold site symmetry, and other atoms are located at site 12d with two-fold site symmetry. The DMI in these alloys stems from their cubic chiral symmetry, and the magnetic phases in these systems resemble those found in B20 materials~\cite{tokunaga2015new, karube2016robust, karube2018disordered, yu2018transformation}. Observations reveal the presence of additional magnetic phases that were not present in the initial B20 materials. 

\begin{figure}[ht!]
\centering
\includegraphics[width=0.4\textwidth]{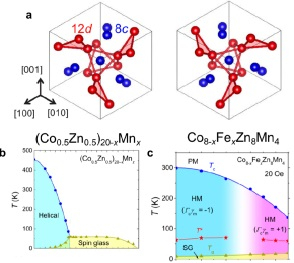}
\caption{(a) Unit cells of the two enantiomers of the Co-Zn-Mn alloys viewed from the [111] axis, (b-c) magnetic phases for (Co$_{0.5}$Zn$_{0.5}$)$_{20-x}$Mn$_x$ and Co$_{8-x}$Fe$_x$Zn$_8$Mn$_4$. Reproduced from reference~\cite{tokunaga2015new}.}
\end{figure}

Large differences in magnetic transition temperature and modulation period can be observed due to the distinctive properties of Co-Zn-Mn alloys, such as their wide solid-solution range and controllable Mn content~\cite{tokunaga2015new, karube2018disordered}. Fe content can be added to the Co site to change the skyrmion helicity, as was seen in the case of Co$_{8-x}$Fe$_x$Zn$_8$Mn$_4$~\cite{karube2018controlling}. Similar to Mn$_{1-x}$Fe$_x$Ge~\cite{shibata2013towards, grigoriev2013chiral}, this change in skyrmion helicity is caused by a constant fluctuation in DM interaction. For the origin of different topological spin textures in these alloys~\cite{karube2016robust, karube2018disordered, karube2017skyrmion, yu2018transformation}, Mn doping is an important factor. It increases the randomness in 12d site occupancies, which leads to the formation of stable skyrmion of larger size because of the weak pinning effect. Therefore, when the skyrmionic to conical phase transition is avoided under field cooling process~\cite{karube2018disordered, karube2017skyrmion}, the equilibrium skyrmion phase pocket located just below T$_c$ can be considered as metastable/supercooled state. When investigating the thermodynamically stable skyrmion lattice in MnSi, Oike et al.~\cite{oike2016interplay} made the first observation of this feature. These Co-Zn-Mn alloys need a moderate cooling rate to achieve a high pinning effect and an enhanced activation energy barrier, even though a sufficiently high cooling rate is necessary to prevent the skyrmion to conical phase transition during the field-cooling process. Several skyrmion materials with interesting technological applications have been found to exhibit metastable skyrmionic states~\cite{munzer2010skyrmion, okamura2016transition, bauer2016history, wild2017entropy, nakajima2017skyrmion, chacon2018observation, halder2018thermodynamic, qian2018new, bannenberg2019multiple, birch2019increased}. 

Magnetic anisotropy can be tuned by controlling the Mn content as the modified helical structure was observed as a sequence of ferromagnetic domains accompanied by domain walls with steep spin rotations in Co$_{10}$Zn$_{10}$. This is happening due to strong in-plane magnetic anisotropy favoring spin to be aligned along the easy plane. The easy-plane anisotropy is reduced with Mn content in these alloys. The change in magnetic anisotropy led to the structural transitions of skyrmion lattices between triangular and squared ones~\cite{karube2016robust} and the emergence of meron-anti meron lattice. Metatable skyrmion state in Co$_{10-x/2}$Zn$_{10-x/2}$Mn$_x$ (x = 4) undergoes reversible transitions
between triangular lattice and a novel square lattice upon varying T and H. The square skyrmion lattice is stable at low temperatures and low magnetic fields as the easy-plane anisotropy is prominent at those temperatures and fields. 

The role of magnetic anisotropy on different structural transitions of skyrmion lattice was estimated in previous theoretical studies~\cite{do2009skyrmions, lin2015skyrmion}. The square meron lattice was observed in Co$_8$Zn$_9$Mn$_3$ via Lorentz TEM measurements along with the skyrmion phase. Triangular skyrmion lattice possesses a metastable state under field cooling process in Co$_8$Zn$_8$Mn$_4$. Finally, we  would like to introduce the 3D disordered skyrmionic states observed in Co$_{10-x/2}$Zn$_{10-x/2}$Mn$_x$ (x = 6) via SANS and Lorentz TEM. These skyrmionic states are governed by the geometrical frustration of antiferromagnetic interactions between Mn moments located at 12d sites. The parent compound $\beta$-Mn does not exhibit long-range magnetic order due to the geometrical frustration~\cite{nakamura1997strong, stewart2002non, paddison2013emergent}. These frustrations can be signified in the form of spin glass behaviors and complex incommensurate antiferromagnetic states in Co-Zn-Mn alloys over a wide range of x[Fig. 4(b)]. 

The frustrated antiferromagnetic correlations result in fluctuations of Co moments via ferromagnetic interaction with Mn. Such spin fluctuations driven by frustration give rise to a thermally equilibrium-disordered skyrmion phase which is independent of T and H. Thus, one needs to exploit another mechanism of the skyrmion formation in DM-interaction-based magnets including magnetic frustration.

\subsection{Cu$_2$OSeO$_3$}
\begin{figure}[ht!]
\includegraphics[width=0.48\textwidth]{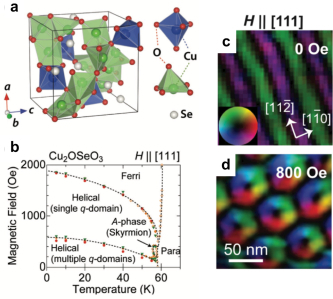}
\caption{(a) Crystal structure of Cu$_2$OSeO$_3$, containing two
inequivalent Cu$^{2+}$ ions with spin-1/2. (b) Magnetic phase diagram of bulk Cu$_2$OSeO$_3$ under magnetic field H along the [111] axis. (c, d) In-plane magnetization distributions of helical structure at H = 0 Oe and SkX at H = 800 Oe in a thin-plate Cu$_2$OSeO$_3$ observed by Lorentz transmission electron microscopy. Reproduced from reference~\cite{seki2012observation}.}
\end{figure}

Another chiral material crystallized into cubic symmetry with space group P2$_1$3 which is the first insulating material hosting nanometer-sized skyrmions~\cite{seki2012observation}. Though this material is identical to B20 materials in a symmetry point of view the atomic arrangements are more complex and entirely different. Cu$^{2+}$ ions with spin-1/2 are positioned at two inequivalent sites and surrounded by different environments [Fig. 5(a)]. Three-fourths Cu(1) sites are surrounded by square pyramides of oxygen ligands and one-fourth at Cu(2) sites are surrounded by trigonal bipyramids. These Cu(1) sites are ferromagnetically coupled to each other and they interact with Cu(2) sites antiferromagnetically. This combination stabilizes a ferrimagnetic ordering in this material below T$_c$ value of 60 K~\cite{kohn1977new, bos2008magnetoelectric}. Lorentz TEM ~\cite{seki2012observation} and neutron diffraction~\cite{adams2012long, seki2012formation} measurements revealed the possible modulation of ferrimagnetic structure up to long-period helical structure stabilized by DM interactions. The smaller T-H phase pocket is suitable for skyrmion lattice state in bulk samples~\cite{adams2012long, seki2012formation}, whereas thin film samples show stable skyrmion states in wide T-H region~\cite{seki2012observation} [Fig. 5(b)]. The skyrmion phase in Cu$_2$OSeO$_3$ at various magnetic fields are shown in Figs. 5(c \& d). The skyrmion states emerged due to the interplay between conduction electrons and topological spin textures, but this can not be anticipated in the case of insulators. The coexistence of magnetic and ferroelectric orders, so-called multiferroicity, is a key feature in producing gigantic magnetoelectric (ME) effects. The cross correlation between magnetism and electricity enables the mutual control of magnetization (polarization) by electric (magnetic) fields. Skyrmions in magnetic insulators can induce electric dipoles or quadruples via SOC, and such multiferroic properties may provide various methods for the manipulation of skyrmions by electric fields. 

\subsection{Inverse Heusler MnPtSn and its alloy}
D$_{2d}$ point group symmetry displayed unique features concerning the DM interaction in high symmetry non-centrosymmetric magnets, and garnered further research on inverse Heusler compound such as Mn-Pt-Sn based Mn$_{1.4}$Pt$_{0.9}$Pd$_{0.1}$Sn. The presence of D$_{2d}$ symmetry enables opposite DM interaction along the x and y axis thus forming opposite helices propagating opposite to each other~\cite{bogdanov1989thermodynamically}. Such D$_{2d}$ magnets are host to antiskyrmion with Q = +1, and their existence was ascertained by theory~\cite{bogdanov1989thermodynamically} and experiment~\cite{nayak2017magnetic} as well. The Lorentz TEM image displays opposite helicities of the antiskyrmion reflecting the anisotropic DM interactions. The vorticity pattern of the antiskyrmion is facilitated by the Bloch-type spin helices along the `x' and `y' axes and N\'eel-type spin helices favor the vorticity pattern along the x$\pm$y axes. Mn$_{1.4}$Pt$_{0.9}$Pd$_{0.1}$Sn thin plates exhibit the antiskyrmion forming the triangular lattice over a wide T-H region, and the antiskyrmion size is about to be 150 nm being reminiscent of magnetic bubbles formed by dipolar interaction. The size of antiskyrmion is film thickness dependent and indicates the importance of dipolar interactions~\cite{malozemoff1979magnetic}. These antiskyrmion textures are governed by anisotropic DM interactions in the underlying D$_{2d}$ lattice symmetry. The conventional dipolar interaction can also lead to the emergence of antiskyrmion textures with Q = -1, which are less affected by anisotropic DM interactions~\cite{malozemoff1979magnetic}. The application of in-plane field to Mn$_{1.4}$Pt$_{0.9}$Pd$_{0.1}$Sn thin plate transits the antiskyrmion into a nontopological spin texture with Q = 0, accompanied by two Bloch lines and topologically equivalent to type II antiskyrmions. The removal of the in-plane magnetic field displays the deformed skyrmions (Q = -1). This affirms the topological transformation from the antiskyrmion (Q = +1) to the skyrmion (Q = -1) via in-field induced nontopological spin texture (Q = 0), which itself a combined shape of half skyrmion (Q = -1/2) and half antiskyrmion (Q = +1/2). The competing anisotropic DM interactions and magnetic dipolar interaction are responsible for the elongated or elliptic skyrmion shape. Thus, It can be stated that dipolar interactions are key players in the formation of skyrmionic or antiskyrmionic spin textures, whereas the anisotropic DM interactions are prominent for the generation of the
antiskyrmion or the elliptically deformed skyrmion in this D$_{2d}$ magnet. 

\subsection{Polar magnets}
This section mainly focuses on the polar materials which lack inversion symmetry and they exhibit DM interactions dominated by their polar symmetry as observed at the interface of heterostructures~\cite{bogdanov2001chiral, xia1997noncollinear, crepieux1998dzyaloshinsky}. These polar magnets show the N\'eel-type skyrmions distinct from those Bloch-type skyrmions discussed for chiral magnets. The polar nature of such magnetic materials does not show any competing magnetic phases other than N\'eel type skyrmions which enhance the stability of such kinds of skyrmions~\cite{kezsmarki2015neel, fujima2017thermodynamically, bordacs2017equilibrium}. This feature of N\'eel-type skyrmions enables them viable candidates for device applications. Now, we would like to introduce a few polar magnetic materials appealing to technological applications.

\subsubsection{GaV$_4$S$_8$ and GaV$_4$Se$_8$:}

GaV$_4$X$_8$ (X = S, Se) are the firstly identified polar achiral ferromagnets exhibiting N\'eel-type skyrmions. These materials fall in the family of lacunar spinels derived from conventional spinel structure AM$_2$X$_4$ upon removal of every second ion positioned at A-site. The crystal structure consists of binary building blocks of (V$_4$X$_4$)$^{+5}$ and (GaX$_4$)$^{-5}$, which are arranged in an alternative manner, thus forming rocksalt-type structure[Fig. 6(a-b)]. The V cluster displays Jahn-Teller distortion in the elongation along one of $\langle$111$\rangle$ crystal axes below T$_s$, which transforms the lattice symmetry to rhombohedral from cubic space group F$\overline{4}$3m~\cite{pocha2000electronic}. (V$_4$X$_4$)$^{+5}$ clusters show long-range magnetic ordering below T$_c$ and it carry the effective spins of S = 1/2. The structural transition temperature T$_s$ and magnetic ordering temperature T$_c$ values are 44 K and 13 K, respectively for GaV$_4$S$_8$~\cite{pocha2000electronic, yadav2008thermodynamic} and their values are 41 K and 18 K, respectively for GaV$_4$Se$_8$~\cite{fujima2017thermodynamically, bichler2011interplay}. The driving factors for the formation of long-period spin spirals and skyrmions in these polar materials are similar to those for chiral magnets. However, the DMI patterns specified to the polar C$_{3v}$ point group, produce different internal spin configurations compared to the chiral magnets. The DM interaction favors cycloidal spin structures with modulated vectors perpendicular to the polar axis, and their spin spiral lies in the plane spanned by its modulation direction and the polar axis. The magnetic propagation vectors q are weakly pinned along one of the three [110] directions perpendicular to the polar axis at zero magnetic field in these aforementioned polar magnets~\cite{kezsmarki2015neel, bordacs2017equilibrium}. A triangular N\'eel-type skyrmion lattice with radial spin texture is found to be stabilized in the form of a superposition of spin cycloids above a critical field value. 

\begin{figure}[ht!]
\includegraphics[width=0.48\textwidth]{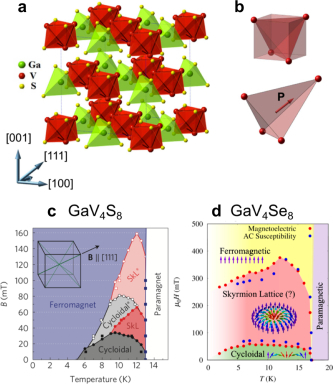}
\caption{(a) Crystal structure of GaV$_4$X$_8$ (X = S or Se), composed of composed of binary building blocks of (V$_4$X$_4$)$^{+5}$ (red hexahedra) and (GaX$_4$)$^{-5}$ (green tetrahedra). (b) Distortion of tetrahedral V-cluster along one of the [111] axes, causing structural phase transition from cubic to polar rhombohedral lattice below T$_s$. (c, d) Magnetic phase diagrams in GaV$_4$S$_8$ and GaV$_4$Se$_8$ . Reproduced from references~\cite{kezsmarki2015neel, fujima2017thermodynamically, butykai2017characteristics}.}
\end{figure}

DM interactions in polar magnets destabilize the conical spin textures modulated along the magnetic field parallel to $\langle 111 \rangle$, which favors the stable N\'eel-type skyrmions in these materials whereas the competing conical phase present in chiral magnets affects the stability of skyrmions and responsible for ultrashort period of modulation of the skyrmionic state. The application of a magnetic field perpendicular to the polar axis leads to the disappearance of the skyrmion-lattice state and there exhibits the stable transverse conical state propagating perpendicular to external magnetic field direction~\cite{bordacs2017equilibrium, leonov2017skyrmion} and its spin spiral plane is deformed as a cone, whose axis is parallel to the external magnetic field. It can be stated that the DM interaction in GaV$_4$X$_8$ strongly favors the magnetic propagations q in the plane perpendicular to the
polar axis. The easy-plane magnetic anisotropy is another factor that favors the skyrmion formation in GaV$_4$Se$_8$ ~\cite{ehlers2016skyrmion, ehlers2016exchange}. Different spin textures identified in these polar magnets are illustrated in Fig. 6(c-d). The limited number of bulk polar materials hosting N\'eel-type skyrmions demand more research in this field to uncover a detailed understanding of the mechanism behind the formation of skyrmion state herein.\\

\subsubsection{VOSe$_2$O$_5$:}
VOSe$_2$O$_5$ is another polar magnet with N\'eel-type
skyrmion state~\cite{kurumaji2017neel, kurumaji2021direct}, and it is composed of VO$_5$ square pyramids having V$^{4+}$ ion with spin S = 1/2 arranged in a square lattice~\cite{meunier1974cristallochimie}. This material is crystallized in tetragonal polar space group P4cc and a point group C$_{4v}$[Fig. 7(a)]. 

\begin{figure}[ht!]
\includegraphics[width=0.48\textwidth]{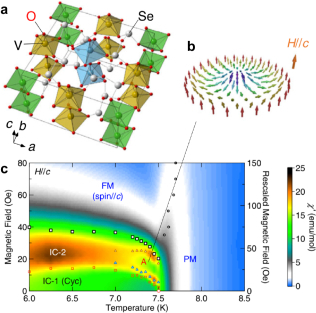}
\caption{(a) Crystal structure of VOSe$_2$O$_5$, composed of VO5 square pyramids (green, blue and yellow polygons). (b, c) Magnetic phase diagram determined by ac magnetic susceptibility $\chi$ measurements. N\'eel-type skyrmions (b) are formed in a triangular lattice at the small phase pocket labeled ``A" in the phase diagram (c). Reproduced from reference~\cite{kurumaji2017neel}.}
\end{figure}

VOSe$_2$O$_5$ exhibits a three-up and one-down type ferrimagnetic transition at T$_c$ = 8K as determined by magnetic susceptibility and SANS measurements under the external magnetic field along polar axis~\cite{trombe1987structure, kim2010experimental}. A cycloidal spin structure appears at zero magnetic fields at T = 5 K, and the enhanced easy-plane anisotropy drives the ferrimagnetic state with spins arranged in the ab plane. Cycloidal spin structures form multidomain states upon modulations along the a- or b-axis and their spin spiral lies in the plane spanned by the polar axis `c' and each modulation direction. Such cycloidal spin modulation was directly observed through polarized SANS. High magnetic fields do not affect magnetic ordering except the spin alignment along the c-axis. Two incommensurate spin states appear in addition to the multi-domain state of spin cycloids below the ferrimagnetic transition field. A triangular N\'eel-type skyrmion lattice is observed in the vicinity of magnetic ordering temperature using extensive SANS measurements~\cite{kurumaji2017neel} while a non-topological but noncoplanar spin structure is referred to as an anisotropic double-q state is proposed for the incommensurate phase-2. Various spin textures observed in VOSe$_2$O$5$ are displayed in Fig. 7(c). Such an anisotropic double-q state has been theoretically predicted in chiral magnets with compass-type anisotropic exchange interactions, which tend to align spins along the bond directions~\cite{chen2016exotic}.

\subsection{Multilayers/Heterostructures}
The realization of skyrmions in nanostructures is essential for the foundation of skyrmion-based spintronic devices~\cite{nagaosa2013topological, fert2013skyrmions, wiesendanger2016nanoscale, finocchio2016magnetic, kang2016skyrmion, garst2017collective, jiang2017skyrmions, fert2017magnetic, everschor2018perspective, zhang2020skyrmion, chen2016exotic}. To achieve this, first one should do either thin-film growth of skyrmion-hosting materials or fabrication of thin-film heterostructures that are compatible with spintronic technology. Thin-film growth is limited by crystal structure control which is still going on~\cite{trabel2017twin, kanazawa2017topological, morikawa2020determination}. It is observed in previous sections that the spin-tectures in non-centrosymmetric magnets are crystal symmetry-dependent. The helicity of Bloch-type or N\'eel-type skyrmion can be altered by crystal chirality, and the coexistence of different chirality domains would lead to different helicity skyrmion lattices within the material. The large-scale growth of thin film with single chirality is necessary to fix this problem which has not been achieved so far. 

\begin{figure*}[ht!]
\includegraphics[width=\textwidth]{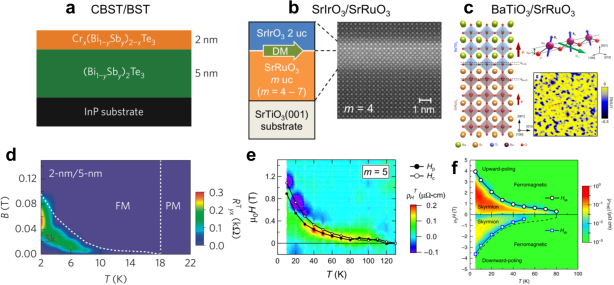}
\caption{Skyrmion formations in various heterostructures consisting of complex compounds. Schematic illustrations of the heterostructures and magnetic phase diagrams for Cr$_x$(Bi$_{1-y}$Sb$_y$)$_{2-x}$Te$_3$/(Bi$_{1-y}$Sb$_y$)$_{2}$Te$_3$ (CBST/BST; panels a, d), SrIrO$_3$/SrRuO$_3$ (SIO/SRO; panels b, e) and BaTiO$_3$/
SrRuO$_3$ (BTO/SRO; panels c, f). Reproduced from references~\cite{yasuda2016geometric, matsuno2016interface, wang2018ferroelectrically}.}
\end{figure*}

Heterostructure interface exhibit N\'eel-type skyrmi-
ons with fixed spin helicity due to fixed polarity of heterostructure. The skyrmion formation at the interface of magnetic metal-heavy metal heterostructure not only showcased the method for formation of N\'eel-type skyrmion but it was proven to be remarkable for potential aplications~\cite{heinze2011spontaneous, romming2013writing}. Additionally, the wide materials choice in interfacial systems offers new avenues of skyrmion science in topological
insulators~\cite{yasuda2016geometric, fijalkowski2020coexistence}, correlated oxides~\cite{vistoli2019giant, ohuchi2018electric, meng2019observation, nakamura2018emergence, matsuno2016interface}, and ferroelectrics~\cite{wang2018ferroelectrically}. Early stages of heterostructures were composed of Fe monolayer~\cite{heinze2011spontaneous} and PdFe bilayer~\cite{romming2013writing, romming2015field} grown on the Ir(111) surface. Fe/Ir interface demonstrated the formation of square skyrmion lattice~\cite{heinze2011spontaneous}, whereas a triangular skyrmion lattice and isolated skyrmions in cycloidal or ferromagnetic state were observed in PdFe/Ir interface~\cite{romming2015field}. DM interactions twist the spin structures in the heterostructures and the lack of inversion symmetry at the interface gives rise to interfacial DM interaction~\cite{bogdanov2001chiral, xia1997noncollinear, crepieux1998dzyaloshinsky, skomski1998unidirectional, fert1980role, smith1976new, zhou2010strength}, which is enhanced by the strong spin-orbit coupling of the heavy-metal layer~\cite{heinze2011spontaneous, belabbes2016hund, yang2015anatomy, yang2018controlling, dupe2014tailoring, khajetoorians2016tailoring}. The local effect of DM interaction mediated by the adjacent magnetic and heavy elements is expressed as the energy gain of $w_D = \vec{d}_{12}\dot (\vec{S}_1 \times \vec{S}_2)$, where $\vec{d}_{12} = d_{12} (\hat{z} \times \vec{u}_{12})$ is DM vector, where $\hat{z}$ and u$_{12}$ are the unit vectors perpendicular to the interface and along the direction connecting from spin site 1 to site 2, respectively. At the micromagnetic level, there is a competition among ferromagnetic interaction, DM interaction and out-of-plane magnetic anisotropy which eventually led to the formation of cycloidal and N\'eel-type skyrmion states as confirmed by spin-polarized scanning tunneling microscopy (SP-STM)~\cite{wiesendanger2016nanoscale, heinze2011spontaneous, romming2013writing}. The size of skyrmions is very small of the order of 1 nm for Fe/Ir interface~\cite{heinze2011spontaneous} and 6-7 nm for PdFe/Ir interface~\cite{romming2013writing, romming2015field}. The standard values of magnetic field and temperature to achieve stabilized skyrmions are far from the industrial application use. Thus, an extensive research effort had been undertaken, which ultimately resulted in stable isolated skyrmion states at room temperature and zero magnetic field~\cite{moreau2016additive, soumyanarayanan2017tunable, woo2016observation, chen2015room, boulle2016room, legrand2017room, yu2016room}. The improvement in interlayer exchange coupling~\cite{chen2015room} and large effective magnetic volume in asymmetric multilayer structures~\cite{moreau2016additive} are the key factors to achieve room temperature stability in skyrmionic states.  The multilayer structure [Ir/Co/Pt]$_{10}$ is constructed for two purposes, one is to enhance the SOC effect to facilitate large DM interaction and the other is to achieve stable magnetic ordering with high magnetic volume against thermal fluctuations. Furthermore, there is a significant reduction in skyrmion size lesser than 100 nm due to two different DM interactions at the Ir/Co and Co/Pt interfaces. Though there are numerous studies available addressing the room temperature stability of skyrmions in multilayer systems~\cite{moreau2016additive, soumyanarayanan2017tunable, woo2016observation, chen2015room, boulle2016room, legrand2017room}, their larger size ranging between 100 nm to 1 $\mu$m is a challenging task. Therefore, the utilization of interfacial DM interactions seems a promising approach to obtain smaller skyrmion sizes in thin film devices. The advancement in first-principles method helped to predict the interfacial DM interaction values and produce several combination of materials~\cite{heinze2011spontaneous, belabbes2016hund, yang2015anatomy, yang2018controlling, dupe2014tailoring, khajetoorians2016tailoring} and further miniaturization of skyrmions might be exploited in higher density memory devices. 

The extensive investigation of mono elemental multi-layer films has shown their promising features like their industrial compatibility, the concept of interfacial DM interaction, suitability to various compound combinations, and many more. The combination of magnetic force microscopy (MFM) measurements and topological Hall effect identified skyrmion formation at the interfaces of magnetic
topological insulators Cr$_x$(Bi$_{1-y}$Sb$_y$)$_{2-x}$Te$_3$/(Bi$_{1-y}$Sb$_y$)$_{2}$Te$_3$ (CBST/BST)~\cite{yasuda2016geometric}, correlated oxides SrIrO$_3$/SrRuO$_3$ (SIO/SRO)~\cite{matsuno2016interface}, and ferroelectrics BaTiO$_3$/SrRuO$_3$ (BTO/SRO)~\cite{wang2018ferroelectrically}[Fig. 8]. The MFM measurements
exhibited dot-like patterns with the opposite out-of-plane
magnetization component to the ferromagnetic background, indicating the skyrmion formation without identification of internal spin arrangements. The observed topological Hall effect signifies the transport mechanism of skyrmion formation. The skyrmion formation in these heterostructures can be coupled to unique materials' properties. The fusion of topological properties in real space and momentum space might bring novel emergent electromagnetic field effects in topological insulators (CBST/BST), while giant responses sensitive to external stimuli may be realized in correlated oxides (SIO/SRO). It is a remarkable instance that the control of skyrmion density and thermodynamic stability can be achieved by manipulating the ferroelectric polarization of BTO/SRO interface~\cite{wang2018ferroelectrically}. The upward-downward polarization results in the shrinking of the skyrmion phase region and a reduction in the magnitude of the topological Hall effect. Therefore, the skyrmion density can be approximated to be proportional to topological Hall effect signal intensity and their stability is largely affected by the ferroelectric polarization in BTO, which can be effective for tuning the degree of inversion-symmetry breaking or DM interaction. Thus, there is a vast scope for materials exploration in thin-film interface systems to manipulate skyrmions for their practical uses and to uncover novel functionalities. 
Till now, we have provided information about skyrmion formation and their stability mechanisms in noncentrosymmetric magnets. The next section will bring light on skyrmion formation in centrosymmetric systems.

\section{Centrosymmetric skyrmions}
The implications of frustrated magnetic systems and the role of inversion symmetry are important to understand before we move on to the discussion of centrosymmetric skyrmions and the current interest in them. Let us consider the example of a single three-atom triangular lattice. Assume that the spins are antiferromagnetically coupled, meaning that they are all trying to align themselves opposite to their neighbor. In this scenario, the system experiences so-called frustration because none of the ensuing spins can align anti-parallel with one another. The spins typically find a compromise by settling at a distance of 120$^{\circ}$ from one another. Let's now, however, enlarge the lattice to a larger triangular shape. Consequently, several different configurations emerge for the system to sustain this 120$^{\circ}$ parameter. This lattice exhibits liquid-like properties at low temperatures because it does not choose a single arrangement but instead keeps alternating. This behavior has led to the designation of this state as spin-liquid. As we have already discussed, skyrmions are typically found in noncentrosymmetric systems, where the DM interaction both causes and stabilizes them. However, more recent studies have looked at novel systems. A characteristic length scale and competing interactions are important prerequisites for yielding the skyrmion state. Trigonal lattices and inversion symmetry are two more ways to satisfy these competing interactions of spin moments in a system. Similar to this, hexagonal magnetic systems have gained interest because of a mediation of the RKKY interaction in conduction electrons and local moments of ``Heisenberg-exchange''. Additionally, researchers have discovered a few new materials to examine as a result of their investigation into these qualities. We will go into more detail about each of these new materials and how they have helped to make centrosymmetric skyrmions a promising new option for realizing the skyrmion state in this section of the review.

\subsection{Gadolinium-based Systems}
In addition to the distinctive characteristics of individual centrosymmetric skyrmions, the more specific case of compounds based on gadolinium offers unique features that highlight the potent capabilities of centrosymmetric skyrmions. Insights into Gd-based compounds, Gd$_2$PdSi$_3$, GdRu$_2$Si$_2$, and Gd$_3$Ru$_4$Al$_{12}$ will be provided in this section.

\begin{figure}
\includegraphics[width=\linewidth]{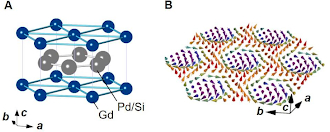}
\raggedright
\caption{} (a) Crystal Structure of Gd$_2$PdSi$_3$, (b) Illustration of the skyrmion phase across Gd$_2$PdSi$_3$ triangular lattice. Reproduced from reference~\cite{kurumaji2019skyrmion}
\end{figure}

Gd$_2$PdSi$_3$ has garnered interest over time because of a wide range of Hall anomalies found in the compound's polycrystalline forms nearly twenty years ago. This form of Gd$_2$PdSi$_3$ demonstrates antiferromagnetic and ferromagnetic correlations in spite of the absence of stable spin-glass anomalies ~\cite{kurumaji2019skyrmion, hirschberger2020topological}. In addition, giant magnetoresistance in its magnetic ordered state and those above the N\'eel temperature (T$_n$) have been observed to exhibit this complexity in negative coefficients of electrical resistivity values. Geometrically speaking, frustrated magnets have competing interactions between neighboring electron spins. Magnetic order suppression happens if these interactions are not satisfied. This leads to fast fluctuations in spin states and the formation of disordered spin states. Spirals and other patterns can frequently be produced by this instability of spin states. The nature of these different states indicates that the magnetic properties of the material can be strongly influenced by even minute modifications in the system. Among these materials is Gd$_2$PdSi$_3$, a skyrmion candidate. The metallic-based magnet Gd$_2$PdSi$_3$ is made up of a complex system of Gd atoms arranged in a triangular lattice throughout the entire hexagonal structure [Fig. 9(a)]. When applying a magnetic field perpendicular to the Gd lattice plane, Gd$_2$PdSi$_3$ exhibits the emergence of a skyrmion lattice state [Fig. 9(b)]. Kurumajo et al.~\cite{kurumaji2019skyrmion} employed x-ray scattering to more clearly define the effects resulting from the individual Gd spins. Recall for a moment that the chemical composition and other physical characteristics of materials, as well as the crustal structure, can be determined with the aid of X-ray scattering techniques. The DM interaction is absent from Gd$_2$PdSi$_3$ because of the AlB$_2$-type crystal structure; instead, the RKKY interaction between the 4f electron moments generates the skyrmionic state. Strong coupling between adjacent conduction electrons and Gd atom moments has been suggested in Gd$_2$PdSi$_3$ through magnetic transitions from field interactions. As a result, non-standard spin structures are found to arrive in the triangular-lattice structure of the Gd atom. It is now possible to observe topological hall effect behavior originating from magnetic skyrmions for the first time in Gd$_2$PdSi$_3$~\cite{kurumaji2019skyrmion}.

\begin{figure}
\includegraphics[width=\linewidth]{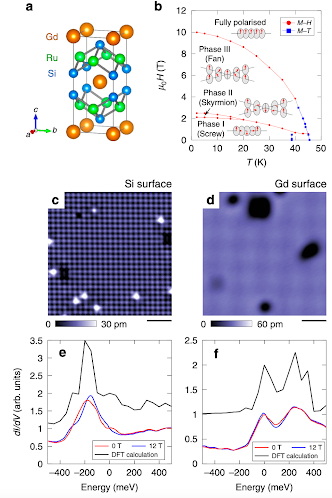}
%\raggedright
\caption{(a) GdRu$_2$Si$_2$ crystal structure. (b) Magnetic phase diagram with magnetic field H applied parallel to the c-axis and schematics of each phase depicted. (c-d) Topographic images of Si-terminated surfaces and Gd-terminated surfaces, respectively. (e-f) Spectral conductance for Si and Gd terminated surfaces. Reproduced from reference~\cite{bouaziz2022fermi}}
\end{figure}
\begin{figure}
\includegraphics[width=\linewidth]{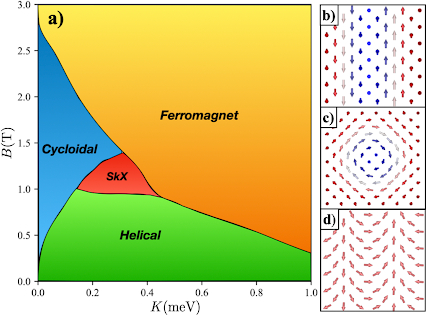}
\caption {} (a) GdRu$_2$Si$_2$ magnetic phase diagram featuring four different phases: Ferromagnet, Cycloidal, Helical, and Skyrmion crystal. (b-d) Schematic of magnetic textures and their z-component values (blue = -1, red = 1). Reproduced from reference~\cite{bouaziz2022fermi}
\end{figure}

The centrosymmetric structure of GdRu$_2$Si$_2$ exhibits an absence of DM interaction (DMI), just like that of its other hexagonal magnetic family of Gd atoms. Conversely, 2D sheets of Gd atoms arranged in a square lattice and scattered with Ru and Si atoms form GdRu$_2$Si$_2$ [Fig. 10(a)]~\cite{bouaziz2022fermi}. Compared to its other Gd-atom counterparts, the unique structure of GdRu$_2$Si$_2$ has smaller lattice constants. Numerous properties result from this smaller lattice constant, including a large topological Nernst effect, a planar hall effect, and a "giant" topological hall effect (GTE)~\cite{bouaziz2022fermi}. Recent research has been able to demonstrate from first principles that materials like GdRu$_2$Si$_2$ and Gd$_2$PdSi$_2$ form single-q skyrmion states through the RKKY interaction. It has been established that the ferromagnetic interaction-generated topological spiral pattern is formed inside the Gd atom plane and is weakly coupled along the x-axis. The application of a magnetic field causes the formation of skyrmion crystals [Fig. 11(a-d)]. Similar relationships have been established for both materials' Fermi surface and formation energy, interactions from mobile electrons, and transition temperatures. In this review section, we will attempt to provide evidence, claiming this material, along with other rare-earth metals, hosts for stable skyrmion states.

We will first concentrate on comprehending the mechanism that leads to the formation of the skyrmion state when examining GdRu$_2$Si$_2$ in particular. Initially, different experiments show that different magnetic orders are currently provided by Gd-atom 4f orbitals when we examine this material and its crystal structure. Very few interactions originate from the \textit{Si-3p} and \textit{Gd-5p} orbitals~\cite{yasui2020imaging}. This type of magnetic ordering also originates from itinerant electrons from the 4d orbitals of the Ru-atoms. As with local magnetic moments, the coupling between itinerant electron states is caused by distinct modulations of itinerant electrons that are connected to magnetic structures. Additionally, it has been confirmed~\cite{khanh2020nanometric} that itinerant electrons mediate four spin interactions in GdRu$_2$Si$_2$. Close examination revealed that there is little evidence of a geometrically frustrated lattice in GdRu$_2$Si$_2$, which forms a double-q skyrmion lattice. As previously mentioned~\cite{yasui2020imaging}, it has also been demonstrated that the application of a magnetic field leads to multiple magnetic ordering.                                 
Now, we focus on defining similar mechanisms that give rise to the skyrmion state for another Gd-based magnet, Gd$_2$Ru$_4$Al$_{12}$. A member of the rare-earth family R$_3$Ru$_4$Al$_{12}$, Gd$_2$Ru$_4$Al$_{12}$ has a distinctive crystallographic structure that has garnered attention over time due to its triangular and Kagome lattice features. Not much research has been done on this particular family despite these intriguing characteristics. However, it was recently discovered~\cite{chandragiri2016magnetic} that the Gd-atom derivatives of the family reveal intriguing magnetic properties. 	                                                                              \begin{figure}
\includegraphics[width=\linewidth]{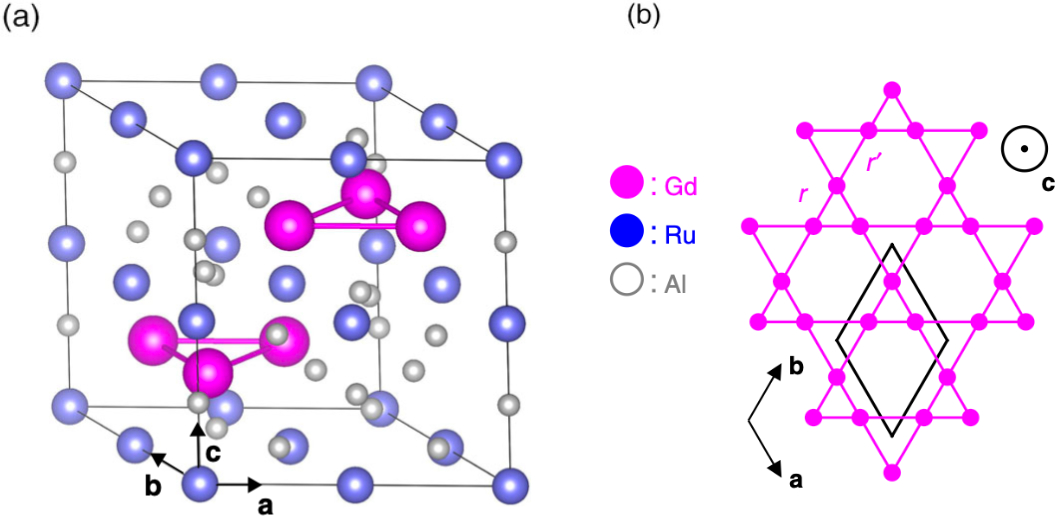}
\raggedright
\caption{} (a) Gd$_3$Ru$_4$Al$_{12}$ Crystal Structure. (b) Distorted Kagome net formed by Gd atoms of alternating distances r and r'.
\end{figure}

\begin{figure*}[!ht]
    \center
    \includegraphics[width=\textwidth]{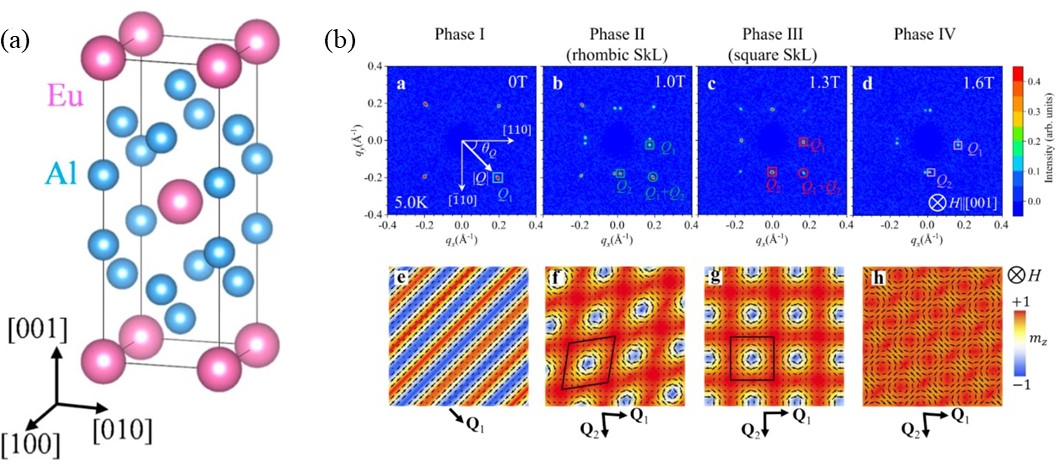}
    \caption{(a) EuAl$_4$ Crystal Structure. (b) Magnetic-field dependence of small angle neutron scattering (SANS) of EuAl$_4$. (e-f) Rhombic skyrmion lattice and square skyrmion lattice. (g) vortex-lattice (h) spin textures. Reproduced from reference~\cite{takagi2022square}}
  \end{figure*}

When Gd$_3$Ru$_4$Al$_{12}$ is first observed, it consists of two layers stacked parallel to the c-axis: Gd$_3$Al$_4$ and Ru$_4$Al$_8$ [Fig. 12(a)]. Once geometrically frustrated magnetism was observed, the Gd-atoms form trihexagonal patterns known as Kagome nets [Fig. 12(b)], which have historically been significant~\cite{wang2021meron}. Larger and smaller triangles with shared vertices and varying edge lengths are presented by Gd$_3$Ru$_4$Al$_{12}$, in place of the usual Kagome-net shape. Geometrically frustrated, Gd$_3$Ru$_4$Al$_{12}$ has competing antiferromagnetic and ferromagnetic states about temperature and H. Findings on this Kagome lattice structure~\cite{swain2022atomic} indicate the emergence of small-scale (2-3 nm) skyrmions through spin-spin interactions. Furthermore, at intermediate field strengths, the magnetic ordering of the system is found to change from a native helical phase to a skyrmion phase. The competition in its crystal structure and its geometric frustration drive Gd$_3$Ru$_4$Al$_{12}$, unlike typical skyrmion candidates. Likewise, it has been demonstrated that the berry curvature of the skyrmion is the source of the skyrmion-driven topological hall effect (STHE) when the system's itinerant electrons are coupled to local moments~\cite{swain2022atomic}. In spintronic applications, a lower energy loss and a much smaller current requirement have been predicted for Gd$_3$Ru$_4$Al$_{12}$ skyrmions at 2.8 nm scale. Furthermore, these smaller-scale skyrmions have been thought to host more stable and long-lasting skyrmion interactions, similar to their other centrosymmetric counterparts~\cite{swain2022atomic}, in contrast to those in the noncentrosymmetric and DMI-driven skyrmions.

\subsection{Oxides and Other Intermetallics}
This section will cover a discussion on other skyrmion hosting materials, EuAl$_4$ and SrFeO$_3$. We will outline the fundamental mechanisms that each one possesses as well as the various experimental techniques and models that are employed to define these attributes. We will provide a detailed description of the distinctions between the materials listed above and their Gd-based substitutes in particular. We aim to provide a definitive review demonstrating that centrosymmetric skyrmions can be formed in a variety of materials by examining the different properties that corroborate the evidence for skyrmion formation. We start with reviewing the material EuAl$_4$, which is centrosymmetric.

EuAl$_4$ is a material exhibiting skyrmion lattice spontaneous rotational symmetry breaking. The material has a square lattice of localized spin Eu ions in a tetragonal structure [Fig. 13(a)]. Seven magnetic phases, including paramagnetic and ferromagnetic phases, can be seen when examining the magnetic field-vs-temperature diagram. In particular, four stages are observed at different temperatures~\cite{gen2023rhombic} when there is no magnetic field [Fig. 13(b) upper panel]. The transition from tetragonal to the orthorhombic structure has been observed using single crystal x-ray diffraction at a temperature of T = 12.2 K. Similarly, it has been demonstrated that the magnetic phases are represented by small angle neutron and resonant x-ray scattering as single-q screw spirals (phases I and V), double-q skyrmion lattices (phases II and III), double-q vortex-antivortex lattices (phases IV and VII), and a double-q meron-anti meron lattice (phase VII)~\cite{gen2023rhombic}. We will try to concentrate on those that are generating the double-q SkL here. Furthermore, it has been discovered that the skyrmion state changes from a square-skyrmion lattice to a rhomboid/rectangular-skyrmion state concurrently with these structural transitions. Research on EuAl$_4$ and its orthorhombic structure aims to clarify the nature of rectangular skyrmion stability~\cite{Hayami_2023}. It is discovered that this transition to the orthorhombic structure of EuAl$_4$ can result in frustrated interactions and the existence of rectangular skyrmion states. The instability of square skyrmion states in EuAl$_4$ increases with increasing magnetic field strength~\cite{Hayami_2023}. The effect of temperature on the system has been identified as fascinating and crucial for further research, but it still requires ongoing examination. It has also been proposed~\cite{takagi2022square} that the transition from square-SkL states to rectangular-SkL states depends delicately on electronic structure, which implies that it can be tunable by applying external stimuli or substituting chemicals. The fact that EuAl$_4$ is the first material made of stacking magnetic `Eu' and nonmagnetic `Al' atoms to form skyrmion spin textures sets it apart from other centrosymmetric skyrmion materials. In the end, this shows that even materials with relatively simple structures, such as EuAl$_4$, can give rise to multiple skyrmionic states, opening up new possibilities for related rare-earth candidates.

\begin{figure}
\includegraphics[width=0.5\textwidth]{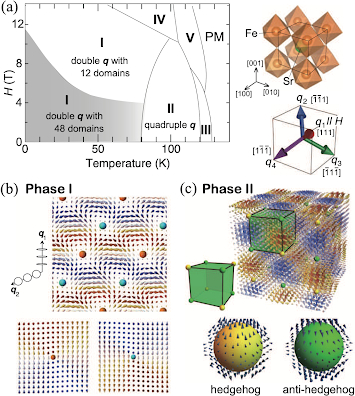}
%\raggedright
\caption {} (a) Crystal structure of SrFeO$_3$ and magnetic phase diagram. (b) Double-q spin structure in Phase I and (c) quadruple-q spin structure in Phase II. Reproduced from reference~\cite{ishiwata2011versatile}
\end{figure}

We now focus on a potential non-Gd-based candidate, SrFeO$_3$, which is thought to present the skyrmion state. Generally, skyrmions are found in noncentrosymmetric materials produced by the DMI, on the other hand, other effects like spin frustration can produce a variety of promising helical spin textures. Higher symmetry lattices preserve free energy degeneracy, these spin textures become extremely intriguing when presented in the cubic lattice. One such material is SrFeO$_3$, which has helical ordering and retains its cubic lattice above transition temperatures [Fig. 14(a)]. Differential helical spin ordering emerges from the competition between ferromagnetic nearest-neighbor and antiferromagnetic next-nearest neighbor interactions~\cite{ishiwata2011versatile}. Further investigation into these distinct helical phases has revealed that a large unconventional Topological Hall Effect can be produced, specifically within SrFeO$_3$ phases I and II, by temperature and the application of an external magnetic field. It has been discovered that the magnetic phases structure of phases I and II are topological, anisotropic double-q, and isotropic quadruple-q helimagnetic structures, respectively, using single crystal neutron diffraction techniques~\cite{ishiwata2011versatile}[Fig. 14(b-c)]. The results of this study show that both structures remain stable in the absence of an external magnetic field. To be more precise, the double-q form illustrates the versatility of centrosymmetric lattice candidates in accommodating distinct spin states. Nevertheless, since the emergent magnetic flux is expected to be directed perpendicular to the external field, the study concluded that additional research is necessary to fully understand the observed Hall resistivity. Likewise, the investigation has discovered that magnetic monopoles emerge during the quadruple-q phase. Although further research is necessary, it is suggested that this variety of spin structures observed in SrFeO$_3$ can be expected to be shared by other similar itinerant high symmetric magnets. 

\section{Applications perspectives of Skyrmions}

\subsection{Topological Hall Eﬀect}
The discussion of any rich physics concepts is inevitable without its technological applications, thus this section will introduce several exotic emergent phenomena and applications owing to skyrmions.
The universal phase diagram exhibits helical structures depending on the temperature, pressure, and magnetic field of the system. Spin-orbit coupling, anisotropy energy, and other relativistic phenomena cause these structures to form~\cite{zadorozhnyi2022theory}. Magnetic moments revolving into helical shapes are known as helical phases. When an external magnetic field is introduced, these helical phases transform into conical phases. At some temperature, called the critical temperature, a stable skyrmion phase arises between the helical and conical phases as a result of further amplification in a magnetic field. The interaction between magnetic and thermal fields and conduction electrons causes the Hall Effect. Hall voltage, the signal detected in the Hall Effect, is dependent on the amount of thickness of the material~\cite{stricker1968hall}. The momentum-space topology of the non-trivial spin texture provides the Berry phase, a non-vanishing Berry phase contributes to a movement of electrons, creating a transverse current. The resulting topological charge density of spin textures causes the topological Hall effect, and the driven current reflects the real-space Berry curvature. When the spin texture of a skyrmion causes it, and called as Skyrmion Hall Effect. The topological Hall effect is not necessary for skyrmion formation as supported by scalar spin chirality for non-coplanar spin structures exhibiting the topological Hall effect, this approach is still actively being used for detecting the skyrmions combined with other experimental techniques.

\subsection{Electronically controlled skyrmion motion}
Many remarkable characteristics of skyrmion dynamics have developed over time, such as the utilization of skyrmion motion in memory devices and the skyrmion Hall effect, which are driven by current. It is commonly known that ferromagnetic domain walls are a primary candidate for spintronic applications utilizing the spin-transfer torque mechanism~\cite{parkin2008magnetic, yamaguchi2004real, tsoi2003magnetic, yamaguchi2004real}. High current densities of the order of 10$^{10}$-10$^{12}$ A/m$^2$~\cite{grollier2003switching, tsoi2003magnetic, yamaguchi2004real, yamanouchi2004current, parkin2008magnetic} are necessary to induce DW motion and generate significant heat in electrically controlled domain walls (DWs). While DWs are more favorable for this application, skyrmions have an advantage since they can be driven by a relatively low current density of 10$^{6}$-10$^{8}$ A/m$^2$~\cite{jonietz2010spin, schulz2012emergent, yu2012skyrmion}. The effect caused by skyrmion dynamics was initially detected experimentally in MnSi~\cite{jonietz2010spin, schulz2012emergent}. Subsequently, comparable characteristics were seen in a multilayer system~\cite{woo2016observation, yu2020motion} and another chiral magnet, FeGe. Moreover, the alignment of skyrmions in a single direction is preferred for their manipulation in electrical devices; yet, the motion of skyrmions is influenced by the skyrmion Hall effect. To mitigate the mutual effects of the topological Hall effect and the skyrmion Hall effect, antiferromagnetic or ferrimagnetic materials were used~\cite{legrand2020room, barker2016static, zhang2016antiferromagnetic, hirata2019vanishing}. The antiskyrmions can also be utilized to drive this effect since, at a specific incident angle of current, their anisotropic in-plane spin textures can generate zero Hall angle.~\cite{huang2017stabilization}. Other phenomena include the spin Hall effect, which is caused by spin accumulation mediated by spin-orbit torque at the interface with the ferromagnetic layer~\cite{woo2016observation, sampaio2013nucleation, jiang2015blowing}. Apart from these discussed phenomena, electric currents can be used to create, delete, and detect the skyrmions.

\subsection{A few other emergent phenomena}
The concurrence of magnetism and ferroelectricity in the same material is a vital parameter for magnetoelectric effects which was attributed to previous findings in perovskites transition metal oxides, BiFeO$_3$~\cite{wang2003epitaxial} and TbMnO$_3$~\cite{kimura2003magnetic}. Various techniques of manipulating skyrmions by electric fields could be produced by the electric dipole or quadrupole caused by skyrmions under spin-orbit coupling, especially in magnetic insulators. The degree of polarization induced by a magnetic field can be used to comprehend multiferroicity in any material. This polarization can be induced along a specific axis/plane depending upon the direction of the magnetic field as observed in the case of Cu$_2$OSeO$_3$ where polarization is induced along [001] and [111] in all the magnetic phases of helical, ferrimagnetic, and skyrmion states when magnetic field is subjected parallel to [110] and [111] directions and no polarization when H $\parallel$ [001]. This directional dependence of magnetoelectric properties can be correlated to crystalline symmetries facilitating the electrical polarization as described for Cu$_2$OSeO$_3$~\cite{tokura2020magnetic} which is further ascribed by spin-dependent hybridization mechanism~\cite{jia2006bond, jia2007microscopic}. The coupling of skyrmions with electric polarization ensures that skyrmions can be controlled by the electric field in the form of their rotation and stability. The measured magnetoelectric effects demonstrated that skyrmions may be controlled with minimal power consumption and minimal heat loss~\cite{okamura2016transition, huang2018situ}.

\section {Conclusion and outlook}
Several skyrmion host compounds based on the DM interactions characteristic feature of noncentrosymmetric systems have been demonstrated in this review article. Additionally, it has been shown that the RKKY-like spin-charge coupling plays a crucial role in stabilizing the skyrmion states and related topological spin textures, most likely in centrosymmetric systems. Within a specific temperature and magnetic field range, it is discovered that the cubic chiral helimagnets display either the relevant hedgehog lattice phase or the skyrmion lattice phase. These chiral skyrmion host materials have demonstrated the existence of a metastable skyrmion phase that is topologically protected at low temperatures and with little to no magnetic field during the quenching process. This attracted further attention to the possibility of using the skyrmion phase in electronic devices by hosting skyrmions on chiral magnets with high T$_c$ values that are even higher than room temperature. Some ferroelectric chiral insulators are also reviewed, which may be of interest due to their multifunctional properties, including ferroelectricity and skyrmion state. While the polar magnets also display such topological spin textures in bulk samples, the N\'eel-type skyrmionic states are mainly found in magnetic layers, which encourages further investigation to generate more combinations. Interesting skyrmion states have been provided by the heterointerfaces made of the strong-SOC layer or topological insulator and two-dimensional transition-metal chalcogenide (2D TMD) ferromagnets~\cite{gong2017discovery, kim2018large}. In the heterointerface of 2D-TMD and topological insulator~\cite{hamamoto2015quantized}, the quantum topological Hall effect induced by skyrmion formation is also realized. A few Gd-based materials as well as oxides and their intermetallics have shown evidence of the presence of small-sized skyrmions of the order of a few nanometers stabilized by RKKY and its higher-order interactions. These materials without DM interactions show stable skyrmion states and are centrosymmetric or frustrated magnets. For the creation of new emergent electromagnetism phenomena, such as the enormous topological Hall effect, a high skyrmion number is necessary for the effective enhancement of the emergent magnetic field effect. Seldom is the RKKY-induced skyrmion lattice observed in materials, but those that do exhibit some special characteristics, like the transition between single-q helical and triple-q skyrmion lattice states, are interesting. In centrosymmetric magnets, the degrees of freedom related to helicity or vorticity comprise two types of transformations. The first one is associated with the helicity transition between Bloch and N\'eel type skyrmionic states and the vorticity transition between skyrmion and antiskyrmion, while the second one results in the creation of helicity multidomain SkX states. Cubic chiral magnet MnGe has been shown to exhibit the coexistence of RKKY and DM interaction. The DM interaction governs the helicity, but the RKKY interaction drives the small skyrmion-string size in the hedgehog lattice of this material. 
Although the materials mentioned earlier in this review have been thoroughly examined, there are still certain unresolved issues in the field of skyrmion science that may be disregarded in light of new developments. A thorough examination of the different magnetic exchange interactions linked to magnetic systems is necessary for smaller skyrmion sizes, as the stability mechanism for larger skyrmion sizes is well understood and supported by DM interactions. The current materials have low current densities to drive skyrmions, which has led to increased research in the skyrmion field. Notably, this has led to smaller skyrmion sizes being hosted by centrosymmetric magnets. In the near future, one of the future perspectives to investigate is the investigation of centrosymmetric magnets hosting skyrmion states stabilized by RKKY interaction combined with higher order magnetic interactions. 

The dynamics of skyrmion driven by magnons in antiferromagnets are revealed by a recent theoretical study~\cite{jin2021magnon}. Magnons can only be polarized in one direction in ferromagnets, but in antiferromagnets, they can be polarized in both directions. Compared to ferromagnetic skyrmions~\cite{barker2016static}, antiferromagnetic skyrmions have much faster magnetic dynamics~\cite{shen2018dynamics, chen2019electric}. It has been found that circularly polarized magnons can control skyrmion Hall motion, and a skyrmion-magnon scattering mechanism could be useful for spintronics in the future. We are aware that in the presence of an external magnetic field, magnons and skyrmions can both be detected using magneto-Raman spectroscopy and Lorentz TEM, respectively. Polarized magnons are discovered to be essential for comprehending the dynamics of skyrmions. The study of skyrmion dynamics using polarized magnons leads us to believe that skyrmion detection could be achieved with magneto-Raman spectroscopy using a polarized magnetic field. Out of this review article, this is another future goal that should be taken into consideration soon. 

\section{Acknowledgment}
This work is supported by the U.S. Department of Energy, Office of Science, Basic Energy Sciences under Award No.\@ DE-SC0022216. This research used resources of the National Energy Research Scientific Computing Center, a DOE Office of Science User Facility supported by the Office of Science of the U.S. Department of Energy under Contract No. DE-AC02-05CH11231. This research at Howard University used Accelerate ACCESS PHYS220127 and PHYS2100073.

\section{References}

\bibliography{bibliography}

\end{document}